\documentclass{cargese}\usepackage{graphicx}
\usepackage{amsfonts}

\let\footnote\savefootnote
\let\footnotetext\savefootnotetext 
\let\footnoterule\savefootnoterule 
\normallatexbib

\begin{document}
\mainmatter
\begin{flushright}INLO-PUB-15/99 \end{flushright}
\vspace{16mm}
\noindent
{\chaptertitlefont  FLAT CONNECTIONS FOR YANG-MILLS THEORIES ON THE 3--TORUS}


\author{Arjan Keurentjes}


\affil{Instituut-Lorentz for theoretical physics, Universiteit Leiden \\
P.O.Box 9506, NL-2300 RA Leiden, The Netherlands}         
%
\email{arjan@lorentz.leidenuniv.nl }
%

\begin{abstract}
We discuss the moduli space\index{moduli space!Yang-Mills on $T^3$} of flat connections of Yang-Mills theories formulated on $T^3 \times \mathbb{R}$, with periodic boundary conditions. When the gauge group is $SO(N \geq 7)$, $G_2$, $F_4$, $E_6$, $E_7$ or $E_8$, the moduli space consists of more than one component. 
\end{abstract}

\section{Introduction}

The vacuum equations for Yang-Mills theories on a 3--torus\index{Yang-Mills theory!on a 3--torus} with periodic boundary conditions, $F_{\mu \nu} = 0$, always allow a trivial solution where the $A_{\mu}$ are constant matrices in the Cartan subalgebra (CSA) of the gauge group $G$. In $N=1$ supersymmetric Yang-Mills theories\index{Yang-Mills theory!supersymmetric}, the solutions of the vacuum equations are of relevance for the computation of the Witten index\index{Witten index} $\textrm{Tr}(-1)^F$ \cite{Wit1}  (which in this case is equal to the number of quantum vacua). With the theory formulated on $T^3 \times \mathbb{R}$, and assuming that all solutions for the gauge field are of the trivial form, Witten found $\textrm{Tr}(-1)^F = r + 1$, with $r$ the rank of $G$ \cite{Wit1}. 

When formulated on $\mathbb{R}^4$, the theory has a $U(1)$ chiral symmetry, broken to $\mathbb{Z}_{2h}$ by instantons (with $h$ the dual Coxeter number of $G$), and further to $\mathbb{Z}_2$ by gluino condensation, leaving $h$ degenerate vacua \cite{Wit1}. Hence, one finds $\textrm{Tr}(-1)^F = h$. Witten's index is assumed to be an invariant, so $h$ should equal $r+1$. This is the case for $SU(N)$ and $Sp(N)$, but not for orthogonal or exceptional gauge groups.

The paradox was recently resolved, when it was shown that theories with orthogonal or exceptional gauge groups on a 3--torus, admit non-trivial solutions to the vacuum equations for the gauge fields \cite{Wit2,Keur1,Keur2,Kac,Borel}. Taking these extra solutions into account one finds \cite{Wit2}
\begin{equation} \label{Cox}
\textrm{Tr}(-)^F = h = \sum_i (r_i + 1)
\end{equation}
where the summation runs over all components of the moduli space, and $r_i$ is the rank of the unbroken gauge group on component $i$. For theories with $G= SU(N)$ or $Sp(N)$, the moduli space consists of only one component of trivial solutions.  

\section{Non-trivial flat connections}

A flat connection on $T^3$ is completely characterised by 3 holonomies\index{holonomy} (Wilson loops\index{Wilson loop}) around the non-trivial cycles of the torus:
\begin{equation}
\Omega_k \ =\ P \exp \left\{ i \int_0^{L_k} A_k(x) dx^k \right\}, \qquad k = 1,2,3.
\end{equation}
Periodic boundary conditions mean that these commute in a simply connected representation of the gauge group $G$. Henceforth we will assume $G$ to be simply connected (as well as simple). Constructing a flat connection is therefore equivalent to finding 3 commuting elements in the group $G$. An obvious way to solve this is to exponentiate 3 elements of the CSA (this corresponds to a trivial flat connection).

A new situation occurs if one picks a group element $\Omega_1$ as holonomy that has a non-simply connected centraliser $C(\Omega_1)$. There are now two options for the remaining two holonomies: when lifted to a simply connected covering $\tilde{C}(\Omega_1)$ of $C(\Omega_1)$, they either commute, or they commute up to a non-trivial element $z$ of the centre of $\tilde{C}(\Omega_1)$, which is a lift of the identity of $C(\Omega_1)$. The first case is, up to conjugation, equivalent to the trivial case. The second case is different. On the subtorus spanned by the 2 and 3-direction, one has a case of 't Hooft's twisted boundary conditions \cite{Hooft} in the gauge group $C(\Omega_1)$. 

The subgroup of $G$ commuting with all three holonomies has a rank that is less than or equal to the rank of $G$. Having found a set of commuting holonomies, one can find a component of the moduli space by deforming around the solution while requiring the holonomies to commute. The rank of the unbroken group is constant over a component of moduli space, and if solutions of different rank are found, the moduli space should consist of different disconnected components. 

\begin{figure}[!h]
\begin{center}
\includegraphics[height=3cm]{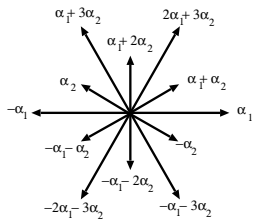}
\end{center}
\caption{the root diagram for $G_2$}
\end{figure}
As an example consider the smallest exceptional group $G_2$. The $g_2$-algebra has an $su(2) \oplus su(2)$ subalgebra, as is clear from the root diagram.
Less immediate from the root diagram is that this is the algebra of an $SO(4) = (SU(2) \times SU(2))/\mathbb{Z}_2$ subgroup of $G_2$ (the $\mathbb{Z}_2$ is diagonal in the $SU(2) \times SU(2)$ covering). This subgroup is non-simply connected, independent of the $G_2$ representation. As a first holonomy one picks the non-trivial centre element of $SO(4)$. This is an element of $G_2$ which has $SO(4)$ as its centraliser. For the remaining two holonomies pick elements in the $SU(2) \times SU(2)$ covering of $SO(4)$ that commute up to the non-trivial element of the diagonal $\mathbb{Z}_2$, and project them to $SO(4)$. This gives 3 commuting elements of $G_2$. The centraliser of these holonomies is a discrete group. Hence this solution is on a different component of the moduli space than the trivial one. It cannot be perturbed, therefore this component of moduli space is an isolated point. All solutions to the vacuum equations are either conjugate to this non-trivial one, or to a trivial one, hence the moduli space of flat $G_2$-connections on $T^3$ consists of 2 components \cite{Keur1}.

\section{Summary of further results}

The results for all simple gauge groups are summarised in a table.
\begin{table}[!h]
\begin{center}
\begin{tabular}{c c c | c c c}\savehline
$G$ & $h$ & $r_i$ & $G$ & $h$ & $r_i$ \\
\savehline
$SU(N)$ & $N$ & $N-1$ & $G_2$ & 4 & 2,0 \\
$Sp(N)$ & $N+1$ & $N$ & $F_4$ & 9 & 4,0,1,0\\
$Spin(2N+1)$ & $2N-1$ & $N$, $N-3$ & $E_6$ & 12 & 6,0,2,0 \\
$Spin(2N)$ & $2N-2$ & $N$, $N-4$ & $E_7$ & 18 & 7,0,1,3,1,0\\
& & & $E_8$ & 30 & 8,0,0,1,2,0,4,0,2,1,0,0 \\
\savehline
\end{tabular}
\caption{Group $G$, dual Coxeter number $h$, rank $r_i$ on $i$-th component}
\end{center}
\end{table}

For orthogonal groups $SO(N \geq 7)$ (or rather the $Spin$-groups, their simply connected coverings) the moduli space also consists of two components, but for $SO(N \geq 9)$ the second component is no longer discrete \cite{Wit2}. For the remaining exceptional groups the moduli space consists of even more components. Via embedding of $G_2$ in these groups a second component of the moduli space can be constructed. $F_4$ and $E_6$ have an $(SU(3)^n)/\mathbb{Z}_3$ subgroup ($n=2,3$ for $F_4$ resp. $E_6$) which allows two non-conjugate non-trivial solutions to be constructed. This brings the number of components for $F_4$ and $E_6$ to 4. The groups $E_7$ and $E_8$ allow still more solutions, as described in \cite{Keur2, Kac, Borel}. In each case equation \ref{Cox} is satisfied. In the table, the components are ordered in a way that exhibits the ``clockwise symmetry'' of \cite{Borel}.

\section{Toroidal compactification}\index{compactification!toroidal}

The results for $T^3$ can be embedded in any $n$-torus with $n \geq 3$. For sufficiently big orthogonal groups, and the exceptional groups $E_8$ also new possibilities occur on higher dimensional tori \cite{Keur2,Kac}. A full classification has not yet been done for these cases. The new solutions, in particular those for the $Spin(32)$ and $E_8$-groups, are of relevance for toroidal compactification of the heterotic and type I string-theories \cite{Wit2,Keur2}. The disconnectedness of the moduli space, and the rank reduction show that they are not contained in the Narain-compactification scheme \cite{Narain}. This is because Narain-compactification assumes Wilson lines with gauge fields taking values in the CSA, and the results presented here show that this is not exhaustive. The models constructed with new solutions for the gauge fields are similar to the CHL-models \cite{CHL}, but allow other possibilities for the reduction of the rank of the gauge group \cite{Keur2}.





%

\end{document}